\documentclass[%
 reprint,
superscriptaddress,
 amsmath,amssymb,
 aps,
]{revtex4-2}

\usepackage{gensymb}
\usepackage{graphicx}
\usepackage{dcolumn}
\usepackage{bm}
\usepackage[dvipsnames]{xcolor}

\begin{document}

\preprint{APS/123-QED}

\title{Ultrafast x-ray scattering reveals composite amplitude collective mode in the Weyl charge density wave material (TaSe\textsubscript{4})\textsubscript{2}I}
\date{\today}

\date{\today}

\author{Quynh L. Nguyen}
\thanks{These authors contributed equally to this work.}
\affiliation{Stanford PULSE Institute, SLAC National Accelerator Laboratory, Menlo Park, California, 94025, United States}
\affiliation{Stanford Institute for Materials and Energy Sciences, SLAC National Accelerator Laboratory, Menlo Park, California, 94025, United States}

\author{Ryan A. Duncan}
\thanks{These authors contributed equally to this work.}
\affiliation{Stanford PULSE Institute, SLAC National Accelerator Laboratory, Menlo Park, California, 94025, United States}
\affiliation{Stanford Institute for Materials and Energy Sciences, SLAC National Accelerator Laboratory, Menlo Park, California, 94025, United States}

\author{Gal Orenstein}
\affiliation{Stanford PULSE Institute, SLAC National Accelerator Laboratory, Menlo Park, California, 94025, United States}
\affiliation{Stanford Institute for Materials and Energy Sciences, SLAC National Accelerator Laboratory, Menlo Park, California, 94025, United States}

\author{Yijing Huang}
\affiliation{Stanford PULSE Institute, SLAC National Accelerator Laboratory, Menlo Park, California, 94025, United States}
\affiliation{Stanford Institute for Materials and Energy Sciences, SLAC National Accelerator Laboratory, Menlo Park, California, 94025, United States}
\affiliation{Department of Applied Physics, Stanford University, Stanford, California, 94305, United States}

\author{Viktor Krapivin}
\affiliation{Stanford PULSE Institute, SLAC National Accelerator Laboratory, Menlo Park, California, 94025, United States}
\affiliation{Stanford Institute for Materials and Energy Sciences, SLAC National Accelerator Laboratory, Menlo Park, California, 94025, United States}
\affiliation{Department of Applied Physics, Stanford University, Stanford, California, 94305, United States}

\author{Gilberto de la Pe\~na}
\affiliation{Stanford PULSE Institute, SLAC National Accelerator Laboratory, Menlo Park, California, 94025, United States}
\affiliation{Stanford Institute for Materials and Energy Sciences, SLAC National Accelerator Laboratory, Menlo Park, California, 94025, United States}

\author{Chance Ornelas-Skarin}
\affiliation{Stanford PULSE Institute, SLAC National Accelerator Laboratory, Menlo Park, California, 94025, United States}
\affiliation{Stanford Institute for Materials and Energy Sciences, SLAC National Accelerator Laboratory, Menlo Park, California, 94025, United States}
\affiliation{Department of Applied Physics, Stanford University, Stanford, California, 94305, United States}

\author{David A. Reis}
\affiliation{Stanford PULSE Institute, SLAC National Accelerator Laboratory, Menlo Park, California, 94025, United States}
\affiliation{Stanford Institute for Materials and Energy Sciences, SLAC National Accelerator Laboratory, Menlo Park, California, 94025, United States}
\affiliation{Department of Applied Physics, Stanford University, Stanford, California, 94305, United States}
\affiliation{Department of Photon Science, Stanford University, Stanford, California, 94305, United States}

\author{Peter Abbamonte}
\affiliation{Department of Physics, University of Illinois at Urbana-Champaign, Urbana, Illinois, 61801, United States}

\author{Simon Bettler}
\affiliation{Department of Physics, University of Illinois at Urbana-Champaign, Urbana, Illinois, 61801, United States}
\affiliation{Materials Research Laboratory, University of Illinois at Urbana-Champaign, Urbana, Illinois, 61801, United States}

\author{Matthieu Chollet}
\affiliation{Linac Coherent Light Source, SLAC National Accelerator Laboratory, Menlo Park, California, 94025, United States}

\author{Matthias C. Hoffmann}
\affiliation{Linac Coherent Light Source, SLAC National Accelerator Laboratory, Menlo Park, California, 94025, United States}

\author{Matthew Hurley}
\affiliation{Department of Physics, Arizona State University, Tempe, Arizona, 85281, United States}

\author{Soyeun Kim}
\affiliation{Department of Physics, University of Illinois at Urbana-Champaign, Urbana, Illinois, 61801, United States}
\affiliation{Materials Research Laboratory, University of Illinois at Urbana-Champaign, Urbana, Illinois, 61801, United States}

\author{Patrick S. Kirchmann}
\affiliation{Stanford Institute for Materials and Energy Sciences, SLAC National Accelerator Laboratory, Menlo Park, California, 94025, United States}

\author{Yuya Kubota}
\affiliation{RIKEN SPring-8 Center, 1-1-1 Kouto, Sayo-cho, Sayo-gun, Hyogo 679-5148, Japan}

\author{Fahad Mahmood}
\affiliation{Department of Physics, University of Illinois at Urbana-Champaign, Urbana, Illinois, 61801, United States}
\affiliation{Materials Research Laboratory, University of Illinois at Urbana-Champaign, Urbana, Illinois, 61801, United States}

\author{Alexander Miller}
\affiliation{Department of Physics, Arizona State University, Tempe, Arizona, 85281, United States}

\author{Taito Osaka}
\affiliation{RIKEN SPring-8 Center, 1-1-1 Kouto, Sayo-cho, Sayo-gun, Hyogo 679-5148, Japan}

\author{Kejian Qu}
\affiliation{Department of Physics, University of Illinois at Urbana-Champaign, Urbana, Illinois, 61801, United States}
\affiliation{Materials Research Laboratory, University of Illinois at Urbana-Champaign, Urbana, Illinois, 61801, United States}

\author{Takahiro Sato}
\affiliation{Linac Coherent Light Source, SLAC National Accelerator Laboratory, Menlo Park, California, 94025, United States}

\author{Daniel P. Shoemaker}
\affiliation{Materials Research Laboratory, University of Illinois at Urbana-Champaign, Urbana, Illinois, 61801, United States}
\affiliation{Department of Materials Science and Engineering, University of Illinois at Urbana-Champaign, Urbana, Illinois, 61801, United States}

\author{Nicholas Sirica}
\affiliation{Center for Integrated Nanotechnologies, Los Alamos National Laboratory, Los Alamos, New Mexico, 87545, United States}

\author{Sanghoon Song}
\affiliation{Linac Coherent Light Source, SLAC National Accelerator Laboratory, Menlo Park, California, 94025, United States}

\author{Jade Stanton}
\affiliation{Department of Physics, Arizona State University, Tempe, Arizona, 85281, United States}

\author{Samuel W. Teitelbaum}
\affiliation{Department of Physics, Arizona State University, Tempe, Arizona, 85281, United States}

\author{Sean E. Tilton}
\affiliation{Department of Physics, Arizona State University, Tempe, Arizona, 85281, United States}

\author{Tadashi Togashi}
\affiliation{RIKEN SPring-8 Center, 1-1-1 Kouto, Sayo-cho, Sayo-gun, Hyogo 679-5148, Japan}
\affiliation{Japan Synchrotron Radiation Research Institute, 1-1-1 Kouto, Sayo-cho, Sayo-gun, Hyogo 679-5198, Japan}

\author{Diling Zhu}
\affiliation{Linac Coherent Light Source, SLAC National Accelerator Laboratory, Menlo Park, California, 94025, United States}

\author{Mariano Trigo}
\affiliation{Stanford PULSE Institute, SLAC National Accelerator Laboratory, Menlo Park, California, 94025, United States}
\affiliation{Stanford Institute for Materials and Energy Sciences, SLAC National Accelerator Laboratory, Menlo Park, California, 94025, United States}

\begin{abstract}
We report ultrafast x-ray scattering experiments of the quasi-1D charge density wave (CDW) material (TaSe\textsubscript{4})\textsubscript{2}I following photoexcitation with femtosecond infrared laser pulses. From the time-dependent diffraction signal at the CDW sidebands we identify an amplitude mode derived primarily from the transverse acoustic component of the CDW static distortion. The dynamics of this acoustic amplitude mode are described well by a model of a displacive excitation, which we interpret as mediated through a coupling to the optical phonon component associated with the tetramerization of the Ta chains.
\end{abstract}

\maketitle


Charge density wave (CDW) materials are low dimensional systems that exhibit spontaneously broken symmetries associated with instabilities in the Fermi surface. These systems are characterized by a modulation of the valence electron density and a corresponding lattice distortion due to the electron-lattice interaction \cite{GrunerBook, Gruner1988}. 
The fundamental collective modes of the CDW ground state correspond to excitations of the amplitude and phase of the condensed electron density and concomitant lattice modulation. 
While frequency-domain inelastic scattering techniques with neutrons \cite{squires_book} or x rays \cite{Schuelke_book,krisch2007inelastic} can generally access collective excitations, they probe the equilibrium spectrum and cannot disentangle couplings among these modes or with other degrees of freedom. Time-domain scattering techniques \cite{trigo2013} can provide a direct probe of these couplings by perturbing and observing different degrees of freedom. Here we use an x-ray free electron laser (XFEL) to study the collective modes of the CDW in (TaSe\textsubscript{4})\textsubscript{2}I (TSI) after ultrafast near-infrared excitation. 
Exploiting the extremely high frequency and wavevector resolution enabled by the time-domain approach, we find that the amplitude mode of this CDW has a strong acoustic character, a manifestation of the composite nature of the CDW.

(TaSe\textsubscript{4})\textsubscript{2}I (TSI) is a quasi-one-dimensional Weyl semimetal that exhibits a CDW instability at a temperature of $T_c = 263$ K \cite{Gressier1982}. As a Weyl-CDW material it has attracted much recent interest \cite{Wang2013, You2016, Shi2021, Gooth2019} due to its potential for realizing a so-called dynamical axion insulator state \cite{Li2010}, a correlated topological phase where the phase mode of the CDW becomes an analogue of the proposed axial field in high energy physics \cite{Peccei1977, Wilczek1987}. Yet, in spite of their relevance to the low-energy electromagnetic response, the CDW collective modes are poorly understood.
Our wavevector-resolved measurements allow us to clearly distinguish the amplitude mode from other lattice modes and to show that the dominant amplitude collective mode is primarily transverse acoustic, a type of distortion which does not typically couple directly to the Fermi surface instability in CDW systems \cite{Lorenzo1998}.

\begin{figure}
    \centering
    \includegraphics{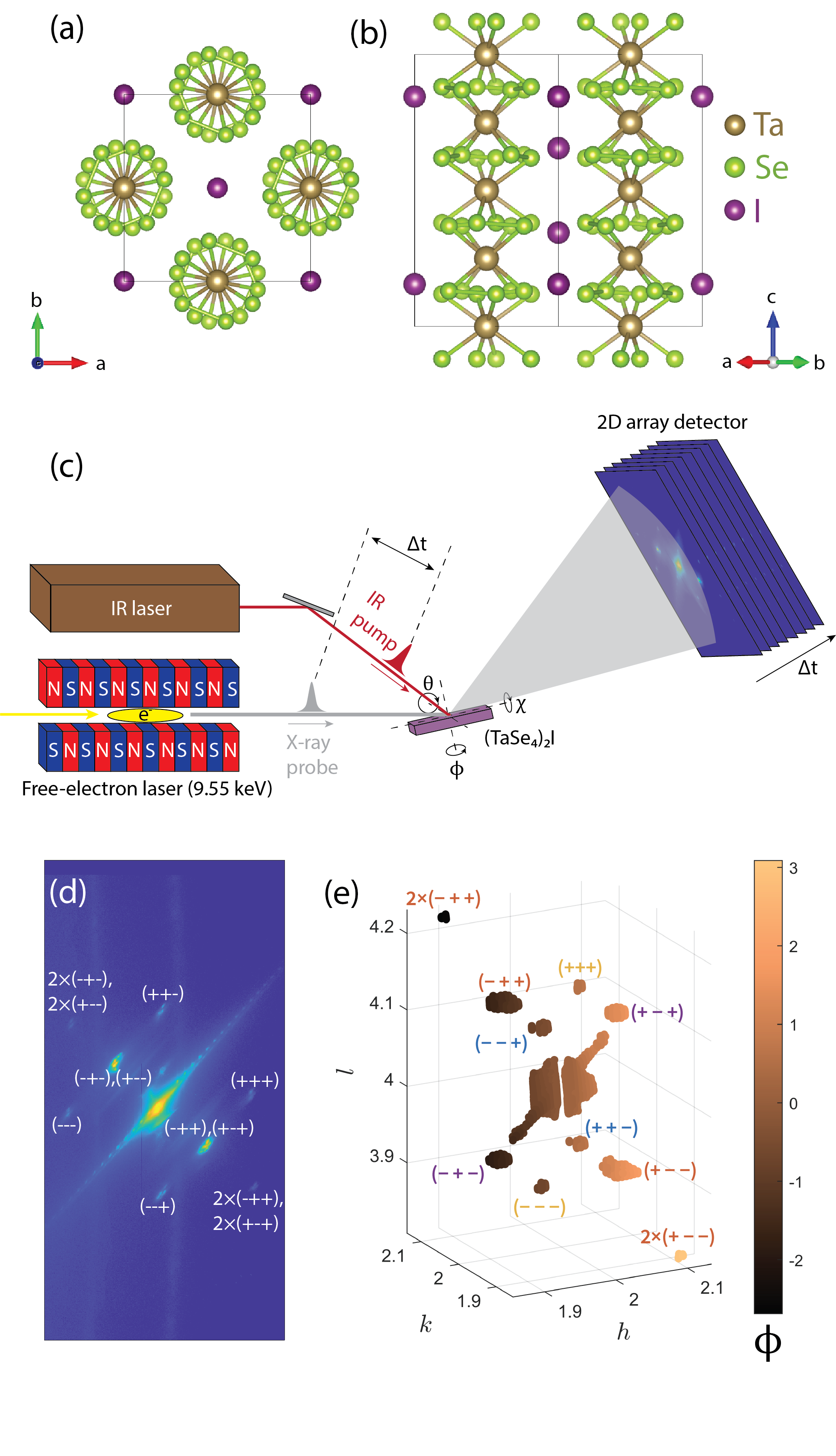}
    \caption{(a,b) Conventional unit cell of TSI in the high-temperature (I422) phase. The lattice constants are $a,b = 9.5317$ \AA{} and $c = 12.761$ \AA. (c) Schematic of the experimental setup. (d) Integrated diffraction image of a scan of the sample azimuth $\phi$ around the (2 2 4) Bragg peak in the CDW phase. Several first- and second-order sidebands are visible. The discrete spots passing diagonally through the Bragg peak are the intersections of the Bragg rod with the Ewald sphere as $\phi$ is varied. (e) Reciprocal space mapping of the scan represented in (d), showing the first-order sidebands as the corners of a box as well as two second-order sidebands. Sidebands whose labels are the same color correspond to the same domain orientation. The colorbar indicates the value of $\phi$ in degrees relative to the (2 2 4) Bragg condition.}
    \label{fig:schematic}
\end{figure}

The high-symmetry tetragonal structure of TSI, with space group I422, is chiral and is comprised of parallel screw-like TaSe\textsubscript{4} chains separated by rows of iodine atoms as shown in Figs. \ref{fig:schematic}(a,b). On account of its chain-like structure TSI has a very high electronic anisotropy \cite{Geserich1986}, with a relatively high conductivity along the chain resulting from a band formed by the Ta $5 d_{z^2}$ orbitals \cite{Gressier1984}. Although the Ta-Ta distances along a chain are all equivalent there are actually two crystallographically and chemically distinct Ta sites due to slightly varying iodine environments, which have different formal oxidation states of +4 and +5. This yields a filling of the Ta $5 d_{z^2}$ band up to the Z-point, which corresponds to a quarter-filling of this band in the approximate one-dimensional Brillouin zone of a single Ta chain along the c-axis \cite{Gressier1984}.
The quasi-1D character of the partially-filled Ta $5 d_{z^2}$ band suggests that the material may exhibit a strong instability to a tetramerization of the Ta atoms along the chain axis through a Peierls-like mechanism. \textit{Ab initio} density functional theory indeed predicts that the I422 phase is unstable to such a distortion \cite{Zhang2020}. However, while the material exhibits a Ta-tetramerization pattern modulated at $\boldsymbol{q}_{\mathrm{CDW}}$, x-ray structural refinement has shown that the dominant component of the CDW distortion below $T_c$ is a frozen transverse acoustic (TA) phonon at $\boldsymbol{q}_{\mathrm{CDW}}$ \cite{vanSmaalen2001}. 
The magnitude of the optical-mode tetramerization is about one-sixth that of the acoustic component, and was only detected using resonant x-ray diffraction to enhanced the sensitivity to the Ta displacements \cite{FavreNicolin2001}.
Understanding the collective modes and their dynamics may illuminate the interplay between multiple distortions which otherwise occur simultaneously in equilibrium.
In the experiment presented here we resolve the amplitude mode of the CDW from other phonon modes, and find that the amplitude mode has a dominant transverse acoustic character. This shows that the amplitude mode of the CDW involves multiple coupled distortions rather than solely the naively expected optical tetramerization mode.

The TSI crystals were grown by chemical vapor transport of the elements in a thermal gradient of 600 {\degree}C to 500 {\degree}C, as described in Refs. \cite{Maki1983, Kim_arxiv}.  The ultrafast x-ray scattering experiments were performed at the XPP instrument of the Linac Coherent Light Source (LCLS) \cite{Emma2010, Chollet2015} at the SLAC National Accelerator Laboratory and at the BL3 hutch of the SPring-8 Angstrom Compact free-electron LAser (SACLA) \cite{Ishikawa2012} at the RIKEN SPring-8 Center. A schematic of the experimental configuration is shown in Fig. \ref{fig:schematic}(c). In the experiments at LCLS (SACLA) a 2 $\mu$m (800 nm) infrared laser pulse with a duration of $\sim$50 fs (45 fs) was used to excite the  TSI sample with a (1 1 0) surface normal at an incidence angle of $7^{\circ}$ from the sample surface. Following excitation the sample was probed with a time-delayed $\sim$30 fs ($<$10 fs) x-ray pulse, and the delay-dependent single-shot diffraction patterns were recorded on a 2D array detector positioned 600 mm (620 mm) away from the sample. The x-ray photon energy and bandwidth were 9.55 keV and 0.5 eV respectively, and the timing jitter between the pump and probe beams was corrected on a single-shot basis to achieve a time resolution of $<80$ fs \cite{Harmand2013, Sato2014}. The pump laser pulses were p-polarized with respect to the sample surface and focused to a spot on the sample with area of $0.15 \times 1.2$ mm\textsuperscript{2} ($0.15 \times 0.2$ mm\textsuperscript{2}). The x-rays were incident at grazing angles between $0.5^{\circ}$-$1^{\circ}$ with respect to the sample surface in order to match the x-ray and optical penetration depths, and the x-ray spot sizes on the sample were $0.1 \times 1.2$ mm\textsuperscript{2} ($0.03 \times 3.4$ mm\textsuperscript{2}). The sample temperature was maintained at $\sim$150 K using a nitrogen gas cryostream cooling system. We do not observe a pump wavelength dependence of the response.

Fig. \ref{fig:schematic}(d) shows the sum of the detector images obtained by scanning the sample azimuthal angle $\phi$ (i.e., the angle parametrizing rotations about the sample normal) near the (2 2 4) crystal Bragg reflection. Several sidebands corresponding to the CDW lattice distortion are visible around the central crystal Bragg peak. Fig. \ref{fig:schematic}(e) shows a mapping of these data into reciprocal lattice indices ($h$ $k$ $l$). The eight first-order sidebands closest to the crystal Bragg peak are at positions ($2\pm\alpha$ $2\pm\alpha$ $4\pm\beta$), where $\alpha \approx 0.055$ and $\beta \approx 0.112$ reciprocal lattice units (r.l.u.)  Also visible are some of the second-order sidebands at ($2\pm2\alpha$ $2\pm2\alpha$ $4\pm2\beta$). 
For simplicity of notation, we will refer to sidebands indices by their sign, e.g.  ($-$ $+$ $-$) represents ($-\alpha$ $\alpha$ $-\beta$), whereas  $2\times $ ($-$ $+$ $-$) is ($-2\alpha$ $2\alpha$ $-2\beta$).
The corresponding labels for each of the sidebands observed are provided in Figs. \ref{fig:schematic}(d,e). Two labels are provided in the cases where two different sidebands fall on the same detector position at different values of $\phi$. According to the structural determination \cite{vanSmaalen2001} the CDW phase features a single wavevector with 4 possible orientational domains, which has been confirmed experimentally by the observation of single-domain x-ray diffraction \cite{FavreNicolin2001}. Thus in Fig. \ref{fig:schematic}(e) the sidebands that lie on a common line passing through the central (2 2 4) crystal Bragg peak belong to the same domain, and sidebands related by 90 deg rotations about the $l$-axis  correspond to different domains.

\begin{figure}
    \centering
    \includegraphics{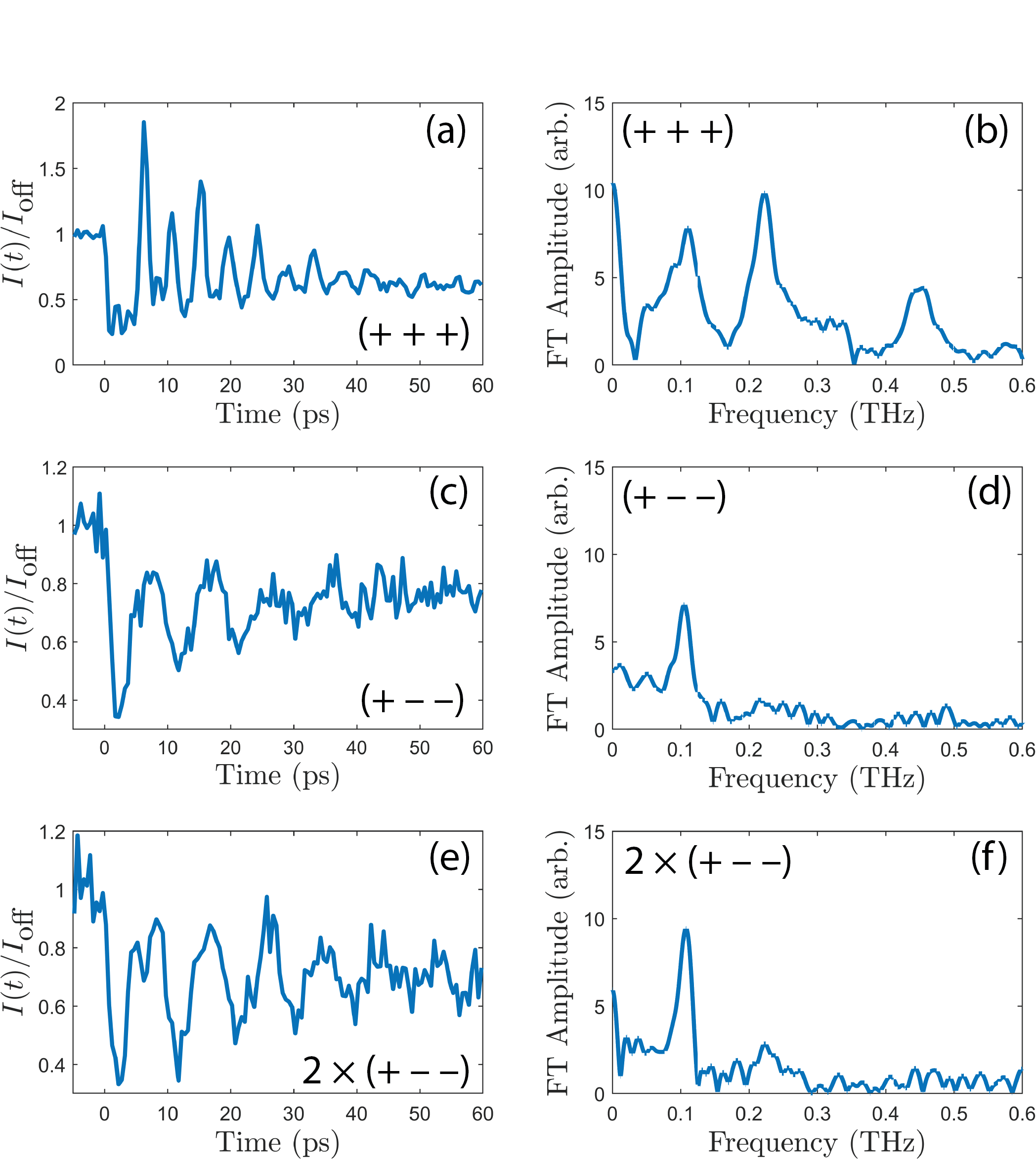}
    \caption{Time-domain signal and Fourier spectra obtained at (a,b) the (2 2 4) ($+$ $+$ $+$) sideband, (c,d) the (2 2 4) ($-$ $+$ $-$) sideband, and (e,f) the (2 2 4) 2$\times$($+$ $-$ $-$) sideband. The measured signals $I$ are normalized by the static sideband diffraction in the absence of the pump laser pulse $I_{\textrm{off}}$ Prominent oscillations at 0.11 THz are observed at approximately the same amplitude for all peaks, whereas the amplitudes of the 0.23 and 0.46 THz modes vary significantly among the different sidebands. Prominent 0.11 THz signal is also observed in the second-order sidebands. These measurements were made during the SACLA experiment.}
    \label{fig:sidebands}
\end{figure}

Figs. \ref{fig:sidebands}(a,c,e) show the normalized diffracted signals $I(t)/I_{\mathrm{off}}$ integrated over regions of interest (ROIs) around the ($+$ $+$ $+$), ($+$ $-$ $-$), and $2 \times$($+$ $-$ $-$) sidebands as a function of pump-probe delay, where $I(t)$ is the recorded x-ray intensity and $I_{\mathrm{off}}$ is the equilibrium intensity. The data obtained at all twelve observed CDW sidebands around the (2 2 4) crystal Bragg peak is provided in the Supplemental Material \cite{SM}. As is evident from the amplitude of the Fourier transforms shown in Figs. \ref{fig:sidebands}(b,d,f), there are three prominent oscillatory components at 0.11, 0.23, and 0.46 THz. The 0.11 THz signal appears at all sidebands with approximately equal normalized amplitude, while the 0.23 and 0.46 THz signals vary significantly with sideband index.
For example, the ($+$ $+$ $+$) sideband in Fig. \ref{fig:sidebands}(a-b) contains all three components, while the ($+$ $-$ $-$) sideband in Fig. \ref{fig:sidebands}(c-d) only shows the 0.11 THz mode. All three components are absent from the I422 crystal Bragg peaks.
Based on the appearance of the 0.11 THz mode in all sidebands with equal relative magnitude we assign this mode to be the amplitude mode of the CDW. Furthermore, this mode modulates both the first- and second-order sidebands with the same 0.11 THz frequency, clearly indicating that it is related to the amplitude mode of the system rather than to an unrelated acoustic phonon which would double its frequency between $\boldsymbol{q}_{\mathrm{CDW}}$ and $2 \boldsymbol{q}_{\mathrm{CDW}}$.
Additionally, the frequency of this mode is close to the transverse acoustic (TA) phonon frequency at $\boldsymbol{q}_{\mathrm{CDW}}$ for the high-temperature phase, which is calculated to be 0.13 THz from measured values of the elastic constants \cite{SaintPaul1996}. 
Importantly, none of the Raman-active  modes soften near the transition \cite{Sugai1985, Schaefer2013}, suggesting that this comparison between similar modes at $T < T_c$ and $T> T_c$ is meaningful.
We further note that the static CDW distortion \cite{vanSmaalen2001} is derived primarily from this TA mode of the high-symmetry structure.

Similar frequencies as those in Fig. \ref{fig:sidebands} have been reported in optical reflectivity measurements \cite{Schaefer2013}. These have been tentatively associated with the CDW phase because they disappear for $T > T_c$. However, optical measurements are only sensitive to long-wavelength excitations and have limited sensitivity to the order parameter, and thus they cannot easily distinguish a collective mode of the CDW from a spectator phonon mode at $\boldsymbol{q}_{\mathrm{CDW}}$ \cite{Schaefer2010,Schaefer2014}. This is especially true if the collective mode does not exhibit softening or any other apparent critical behavior, as is the present case \cite{Schaefer2013}. On the other hand, in our x-ray measurement we are effectively able to determine the polarizations of the observed modes at $\boldsymbol{q}_{\mathrm{CDW}}$ relative to the the CDW distortion. Furthermore, the observation of the 0.11 THz mode equally at all sidebands and also at $2\boldsymbol{q}_{\mathrm{CDW}}$ allows us to unequivocally identify this modulation as originating from an amplitude mode of the CDW.

The other two acoustic modes at $\boldsymbol{q}_{\mathrm{CDW}}$, as determined from calculations based on experimental values of the I422-phase elastic constants \cite{SaintPaul1996}, are a quasi-transverse acoustic (QTA) mode at a frequency of 0.20 THz and a quasi-longitudinal acoustic (QLA) mode at a frequency of 0.38 THz. The calculated frequency of the QTA mode is close to that of the observed 0.23 THz component, and the calculated polarization of the QTA mode is also consistent with the observed variation of the 0.23 THz mode amplitude among the different CDW sidebands \cite{SM}. We thus assign the 0.23 THz component to the QTA acoustic mode at $\boldsymbol{q}_{\mathrm{CDW}}$. Given that the 0.46 THz component appears only when the 0.23 THz component is prominent, and that the maxima of the former are found to temporally coincide with the extrema of the latter, it is likely that the 0.46 THz component is a second harmonic of the QTA  due to nonlinearity of the diffracted intensity with the mode amplitude \cite{Overhauser1971}. This is supported by the fact that no component with a frequency similar to 0.46 THz was observed in the optical transient reflectivity measurements \cite{Schaefer2013}.

The main distortion associated with the CDW seems to have acoustic character, however x-ray diffraction refinement reported a small optical component \cite{vanSmaalen2001}. Furthermore, ultrafast optical experiments \cite{Schaefer2013} and Raman scattering \cite{Sugai1985} report a prominent mode at $\sim 2.7$~THz, which disappears at $T > T_c$ and is ascribed to the collective mode corresponding to this optical component of the CDW. We note that the structure factor of (2 2 4) is not sensitive to this type of distortion \cite{Lorenzo1998} and thus we do not expect this mode in the data in Fig. \ref{fig:sidebands}. The optical component of the CDW, associated with Ta motion, is extremely small but its signature can be enhanced by resonant diffraction on the Ta edge \cite{FavreNicolin2001}. Furthermore, density functional theory (DFT) calculations \cite{Zhang2020} predict an instability of an optical mode of the I422 phase with B\textsubscript{1} and/or B\textsubscript{2} symmetry that would produce a tetramerization of the one-dimensional Ta chain similar to this optical component of the CDW distortion. All these facts taken together suggest that the CDW transition should be thought of as two combined distortions, one optical and one acoustical \cite{Lorenzo1998}.

\begin{figure}
    \centering
    \includegraphics{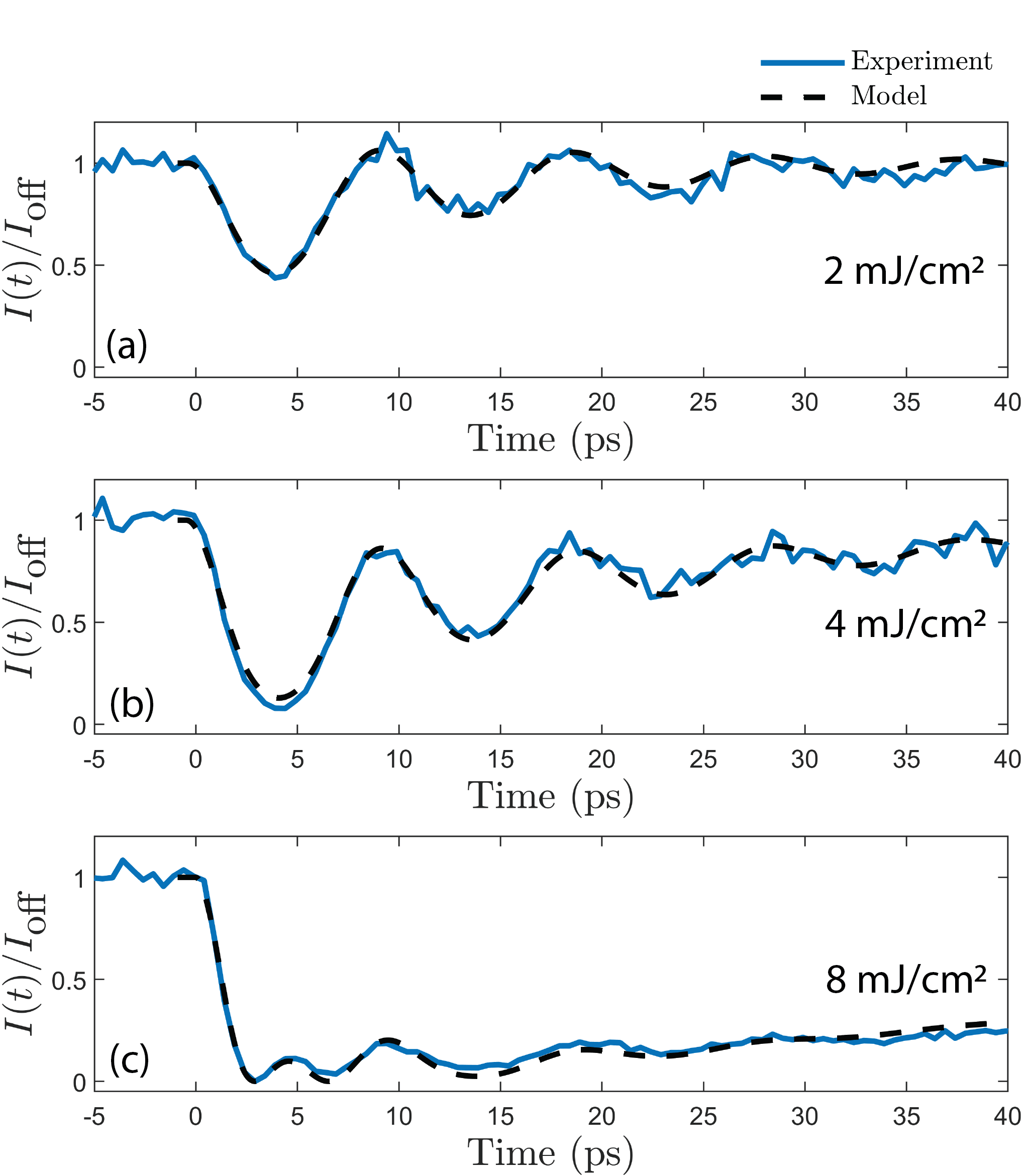}
    \caption{Normalized signal at the (2 2 4) ($-$ $+$ $-$) sideband for pump fluences of (a) 2 mJ/cm\textsuperscript{2}, (b) 4 mJ/cm\textsuperscript{2}, and (c) 8 mJ/cm\textsuperscript{2}, with a pump wavelength of 2 $\mu$m. Dashed lines are fits to the model. These measurements were made during the LCLS experiment.}
    \label{fig:fluence}
\end{figure}

We model the signals at the (2 2 4) sidebands as a displacive excitation of the harmonic oscillator \cite{Zeiger1992} associated with the acoustic component of the CDW \cite{SM}. We define the amplitude of the CDW relative to the equilibrium distortion as $1 + \xi$, with  $ \xi$ the mode displacement, which oscillates around a displaced equilibrium position at $t>0$. We consider a potential of the form

\begin{equation}
    \begin{aligned}
        V(\xi,t) &= \frac{1}{2}a[\xi - \xi_0(t)]^2 \\
        \xi_0(t) &\equiv - \eta \Theta(t)e^{-t/\tau}
    \end{aligned}
\end{equation}

which yields an equation of motion given by

\begin{equation}\label{eq:eom_main}
    \frac{1}{c} \ddot{\xi} = -\xi + \xi_0(t) - \gamma\dot{\xi}.
\end{equation}

The diffracted intensity at $\boldsymbol{q}_{\mathrm{CDW}}$ is proportional to $(1+\xi)^2$ \cite{Overhauser1971}. 
Fig. \ref{fig:fluence} shows the normalized signals observed at the ($-$ $+$ $-$) sideband for a range of excitation fluences. Dashed black lines are the fits obtained for $(1 + \xi)^2$ by solving Eq. \ref{eq:eom_main} for $\xi$ (details provided in the Supplemental Material). We observe no dependence of the mode frequency on the excitation fluence.
The agreement shown in Fig. \ref{fig:fluence} suggests that the acoustic amplitude mode is excited through a sudden displacement of its equilibrium position (the cosine phase of the oscillation is a signature of this effect \cite{Zeiger1992}), although this mode is not itself expected to couple directly to the photoexcited valence electrons \cite{Lorenzo1998}. We argue next that this TA amplitude mode is excited indirectly through its coupling to the optical mode related to the Ta-tetramerization, rather than directly by the photoexcited charge. 

The structural instability of the I422 phase, which is related to the Ta-tetramerization distortion, is derived from a linear combination of the lowest-frequency B\textsubscript{1} and B\textsubscript{2} modes \cite{Zhang2020}. These optical modes interact strongly with the charge and through phonon-phonon coupling can induce a displacive excitation of the acoustic mode. In fact, a phenomenological Ginzburg-Landau model of coupled optical and acoustic modes was proposed to explain several features of the CDW in TSI \cite{Lorenzo1998}. In this model, two soft optical modes of B\textsubscript{1} and B\textsubscript{2} symmetries with quartic bare potentials are coupled to the lattice strain \cite{GLnote}. The inclusion of coupling terms that involve the gradients of the phonon amplitudes yields a CDW lattice modulation at finite wavevector with mixed acoustic and optical character, qualitatively matching the observed distortion \cite{vanSmaalen2001}.
The prominent 2.7 THz mode observed in ultrafast reflectivity and Raman scattering below $T_c$ \cite{Schaefer2013,Sugai1985} has been associated with the optical vibration corresponding to the Ta-tetramerization of the CDW \cite{Schaefer2013}. However, the dominant motion of the amplitude mode observed in our experiments is of acoustic character, and our results shown here further indicate that the order parameter is composed of multiple coupled modes of the high symmetry structure.
Our results also highlight the ability of ultrafast experiments to disentangle complex structural pathways based on their fluence and time dependences.

In conclusion, we present an ultrafast x-ray study of the low energy lattice modes of (TaSe\textsubscript{4})\textsubscript{2}I in the CDW phase. The high resolution in both frequency and wavevector of our measurements performed over multiple CDW sidebands allows us to clearly distinguish lattice modes from the dominant amplitude mode at $\boldsymbol{q}_{\mathrm{CDW}}$. 
Based on the structure factor sensitivity of the different CDW sidebands, we identify a mode with a frequency of $0.11$ THz as an amplitude mode derived from the dominant component of the CDW distortion. This mode has transverse acoustic character, and we observe no measurable softening with pump fluence even for excitations at which the lattice reaches the high-symmetry phase. These two facts suggest that this mode may be excited through coupling to the Ta-tetramerization distortion, which is known to interact strongly with the charge, and that the order parameter and the free energy may involve multiple lattice distortions.

\begin{acknowledgments}
This work was primarily supported as part of the Quantum Sensing and Quantum Materials, an Energy Frontier Research Center funded by the U.S. Department of Energy (DOE), Office of Science, Basic Energy Sciences (BES) at the University of Illinois at Urbana-Champaign under Award No.DE-SC0021238 and at SLAC National Accelerator Laboratory. 
F.M. acknowledges support from the EPiQS program of the Gordon and Betty Moore Foundation, Grant GBMF11069. 
R.A.D. acknowledges support through the Bloch Postdoctoral Fellowship in Quantum Science and Engineering from the Stanford University Quantum Fundamentals, Architectures, and Machines initiative (Q-FARM), and the Marvin Chodorow Postdoctoral Fellowship from the Stanford University Department of Applied Physics. S.B. acknowledges support through the Early Postdoc Mobility Fellowship from the Swiss National Science Foundation (Grant number P2EZP2 191885). Use of the LCLS is supported by the US Department of Energy, Office of Science, Office of Basic Energy Sciences under Contract No. DE-AC02-76SF00515. X-ray measurements at BL3 of SACLA were performed with the approval of the Japan Synchrotron Radiation Research Institute (JASRI) (Proposal Nos. 2021A8037).

\end{acknowledgments}

\providecommand{\noopsort}[1]{}\providecommand{\singleletter}[1]{#1}%

\end{document}


\title{Ultrafast x-ray scattering reveals composite amplitude collective mode in the Weyl charge density wave material (TaSe\textsubscript{4})\textsubscript{2}I: Supplemental Material}

\date{\today}

\author{Quynh L. Nguyen}
\author{Ryan A. Duncan}
\author{Gal Orenstein}
\author{Yijing Huang}
\author{Viktor Krapivin}
\author{Gilberto de la Pe\~na}
\author{Chance Ornelas-Skarin}
\author{David A. Reis}
\author{Peter Abbamonte}
\author{Simon Bettler}
\author{Matthieu Chollet}
\author{Matthias C. Hoffmann}
\author{Matthew Hurley}
\author{Soyeun Kim}
\author{Patrick S. Kirchmann}
\author{Yuya Kubota}
\author{Fahad Mahmood}
\author{Alexander Miller}
\author{Taito Osaka}
\author{Kejian Qu}
\author{Takahiro Sato}
\author{Daniel P. Shoemaker}
\author{Nicholas Sirica}
\author{Sanghoon Song}
\author{Jade Stanton}
\author{Samuel W. Teitelbaum}
\author{Sean E. Tilton}
\author{Tadashi Togashi}
\author{Diling Zhu}
\author{Mariano Trigo}

\maketitle

\section*{Modeling the excitation of the acoustic amplitude mode}\label{ap:DECP}

The time-dependent potential used to model the excitation of the 0.11 THz amplitude mode shown in Fig. 3 is given by

\begin{equation}\label{eq:displacive}
    \begin{aligned}
        V(\xi,t) &= \frac{1}{2}a[\xi - \xi_0(t)]^2 \\
        \xi_0(t) &\equiv - \eta \Theta(t)e^{-t/\tau}
    \end{aligned}
\end{equation}
    
where $\xi_0(t)$ is the location of the potential minimum and $\Theta(t)$ is the Heaviside step function. For $t < 0$ the potential minimum is at $\xi_0 = 0$, and at $t = 0$ it instantaneously shifts by a magnitude given by $\eta$ and gradually relaxes back with a timescale given by $\tau$. The equation of motion for $\xi$ is
    
\begin{equation}\label{eq:disp_eom}
    \frac{1}{c} \ddot{\xi} = -\xi + \xi_0(t) - \gamma\dot{\xi}
\end{equation}
    
where $\gamma$ is the damping, $c \equiv a/m$, and $m$ is the mass.
    
We now motivate the use of a displacive excitation as a model for our experiment. As mentioned in the main text the CDW modulation contains two components, a transverse acoustic distortion and Ta-tetramerization derived from optical phonons \cite{vanSmaalen2001, FavreNicolin2001}. The optical component of the distortion is found to be involved in a predicted structural instability by DFT calculations \cite{Zhang2020}. We thus begin by considering a heuristic Ginzburg-Landau theory  of an anharmonic optical mode with displacement $\xi_1$ in a  quartic potential, coupled to a harmonic acoustic mode with displacement $\xi_2$. The resulting equilibrium CDW lattice distortion will be a linear combination of these two coordinates. The potential expressed to lowest order in mode coordinates is given by
    
\begin{equation}\label{eq:GL_potential}
    V(\xi_1, \xi_2, t) = \frac{1}{2}a_1r(t)\xi_1^2 + \frac{1}{4}b_1\xi_1^4 + \frac{1}{2} a_2\xi_2^2 + g\xi_1\xi_2
\end{equation}

where $a_1, a_2, b_1 > 0$ and we include the effect of the pump through the time-dependent term $r(t)$ \cite{Trigo2019,Schaefer2014}. Here $r(t) = -1$ for $t < 0$, at $t = 0$ it suddenly jumps to a higher value, and for $t > 0$ it gradually and monotonically decays to $r(t \rightarrow \infty) = -1$. The pre-pump equilibrium values for this potential are given by

\begin{equation}\label{eq:equilibrium_vals}
    \begin{aligned}
    \xi_1^{(0)}(t<0) &= \pm \sqrt{\frac{g^2 + a_1a_2}{b_1a_2}} \\
    \xi_2^{(0)}(t<0) &= -\frac{g}{a_2}\xi_1^{(0)}.
    \end{aligned}
\end{equation}

Solving Eq. \ref{eq:GL_potential} for the equations of motion in the $\xi_i$ and normalizing by the equilibrium values given in Eq. \ref{eq:equilibrium_vals} we obtain

\begin{equation}\label{eq:GL_eom}
\begin{aligned}
    \frac{1}{c_1}\ddot{\tilde{\xi}}_1 = -r(t)&\tilde{\xi}_1 - (\Lambda + 1)\tilde{\xi}_1^3 + \Lambda\tilde{\xi}_2 - \gamma_1\dot{\tilde{\xi_1}} \\
    \frac{1}{c_2}\ddot{\tilde{\xi}}_2 &= -\tilde{\xi}_2 + \tilde{\xi}_1 - \gamma_2\dot{\tilde{\xi}}_2
\end{aligned}
\end{equation}

where $\tilde{\xi}_i(t) \equiv \xi_i(t)/[\xi_i^{(0)}(t<0)]$, $c_i \equiv a_i/m_i$, $\Lambda \equiv g^2/a_1a_2$, and $m_i$ are the mass terms. The mode $\tilde{\xi}_1$ directly experiences the effect of the pump through the term $r(t)$, and the subsequent motion of $\tilde{\xi}_1$ then drives the mode $\tilde{\xi}_2$. In our experiment the high-frequency optical component of the CDW corresponds to $\tilde{\xi}_1$ , while the harmonic low-frequency acoustic component corresponds to $\tilde{\xi}_2$. The sudden change in the quartic potential at $t = 0$ for $\tilde{\xi}_1$ caused by the $r(t)$ pump term in Eq. \ref{eq:GL_potential} will induce oscillations of this mode around the slowly recovering potential minimum with a frequency of about 2.7 THz. The high frequency of these oscillations is well outside the bandwidth defined by the values of $c$ and $\gamma$ obtained from fitting the data shown in Fig. 3 to Eq. \ref{eq:disp_eom} (see Table \ref{tab:parameters}). As such, $\tilde{\xi}_2$ will not respond to these fast transients and they can be neglected in the equation of motion for $\tilde{\xi}_2$. If additionally the coupling term $g \ll a_1, a_2$ so that $\Lambda \ll 1$ then the $\tilde{\xi}_2$ term in the equation of motion for $\tilde{\xi}_1$ is negligible. With these assumptions the effective driving in the equation of motion for $\tilde{\xi}_2$ by the $\tilde{\xi}_1$ term can then be approximated by the value of $\tilde{\xi_1}$ that minimizes the uncoupled potential, which is $\tilde{\xi}_1^{(0)}(t) = \sqrt{-r(t)}$. If we then define $\xi_0(t) = \tilde{\xi}_1^{(0)}(t) - 1 = \sqrt{-r(t)} - 1$, $c = c_2$, $\gamma = \gamma_2$, and allow $r(t)$ to have the phenomenologically reasonable form $r(t) \equiv -[1 - \eta\Theta(t)e^{-t/\tau}]^2$, the equation of motion for $\tilde{\xi}_2$ in Eq. \ref{eq:GL_eom} becomes identical to Eq. \ref{eq:disp_eom}.

\begin{table}
\begin{tabular}{|c| c c c|}
\hline
& 2 mJ/cm\textsuperscript{2} & 4 mJ/cm\textsuperscript{2} & 8 mJ/cm\textsuperscript{2} \\
\hline
$\nu_0$ (THz) & 0.106 & 0.105 & 0.109  \\
$\Gamma$ (THz) & 0.033 & 0.035 & 0.069 \\
$\eta$ & 0.24 & 0.43 & 0.91 \\
$\tau$ (ps) & 9.3 & 20.9 & 58.1 \\
\hline
\end{tabular}
\caption{Parameters obtained by fitting solutions of Eq. \ref{eq:disp_eom} (for $(1 + \xi)^2$) to the fluence-dependent data shown in Fig. 3.}
\label{tab:parameters}
\end{table}

The values obtained for the restoring frequency $\nu_0 \equiv \sqrt{c}/2\pi$, damping rate $\Gamma \equiv \gamma c/2$, fluence $\eta$, and relaxation time $\tau$ corresponding to the fitted dashed black lines in Figs. 3(a-c) are given in Table \ref{tab:parameters}.

\section*{Data from all observed CDW sidebands}

\begin{figure}
    \centering
    \includegraphics{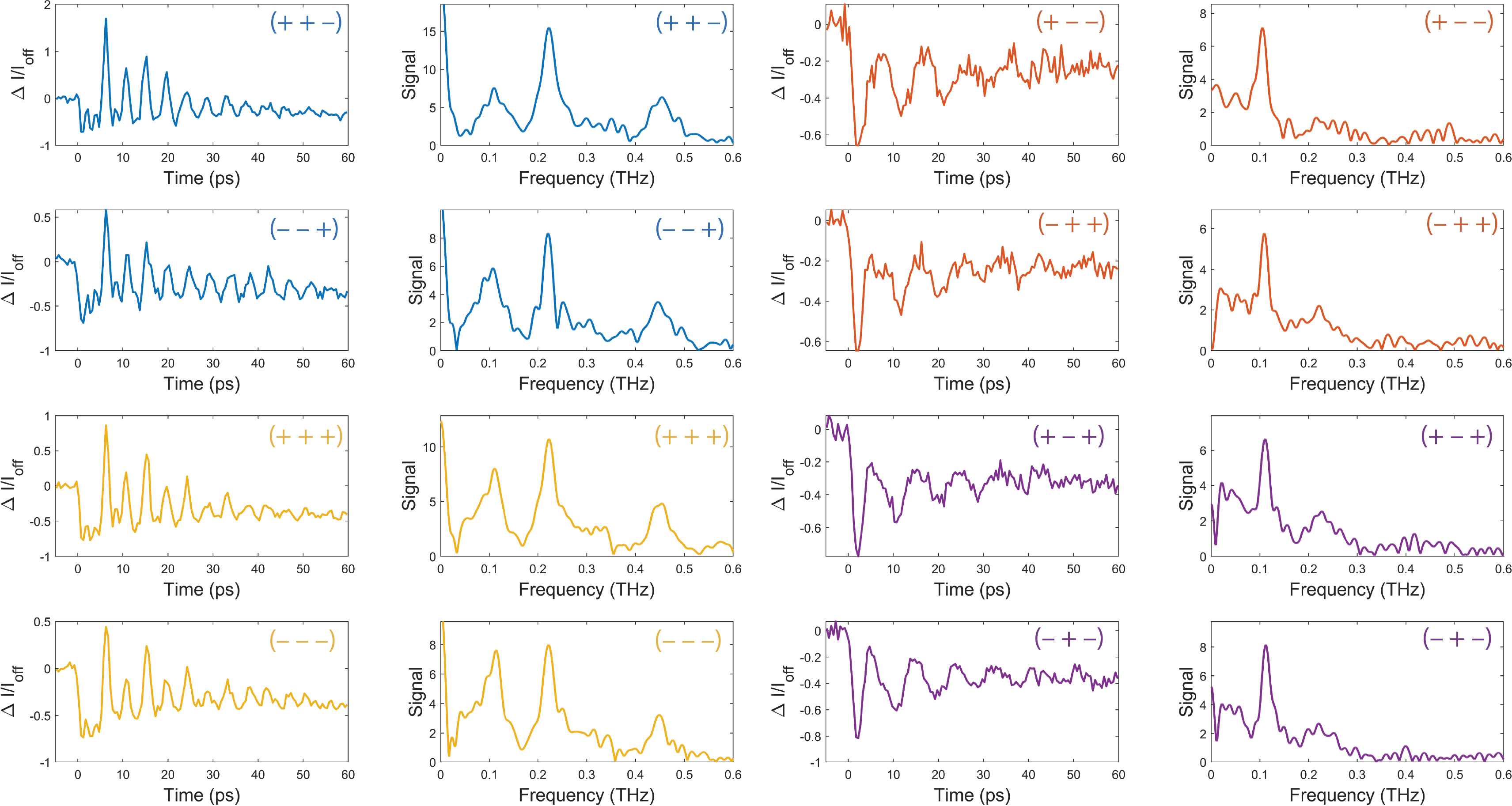}
    \caption{Data from all eight first-order sidebands around the (2 2 4) crystal Bragg peak of TSI.}
    \label{fig:firstorder_all}
\end{figure}

\begin{figure}
    \centering
    \includegraphics{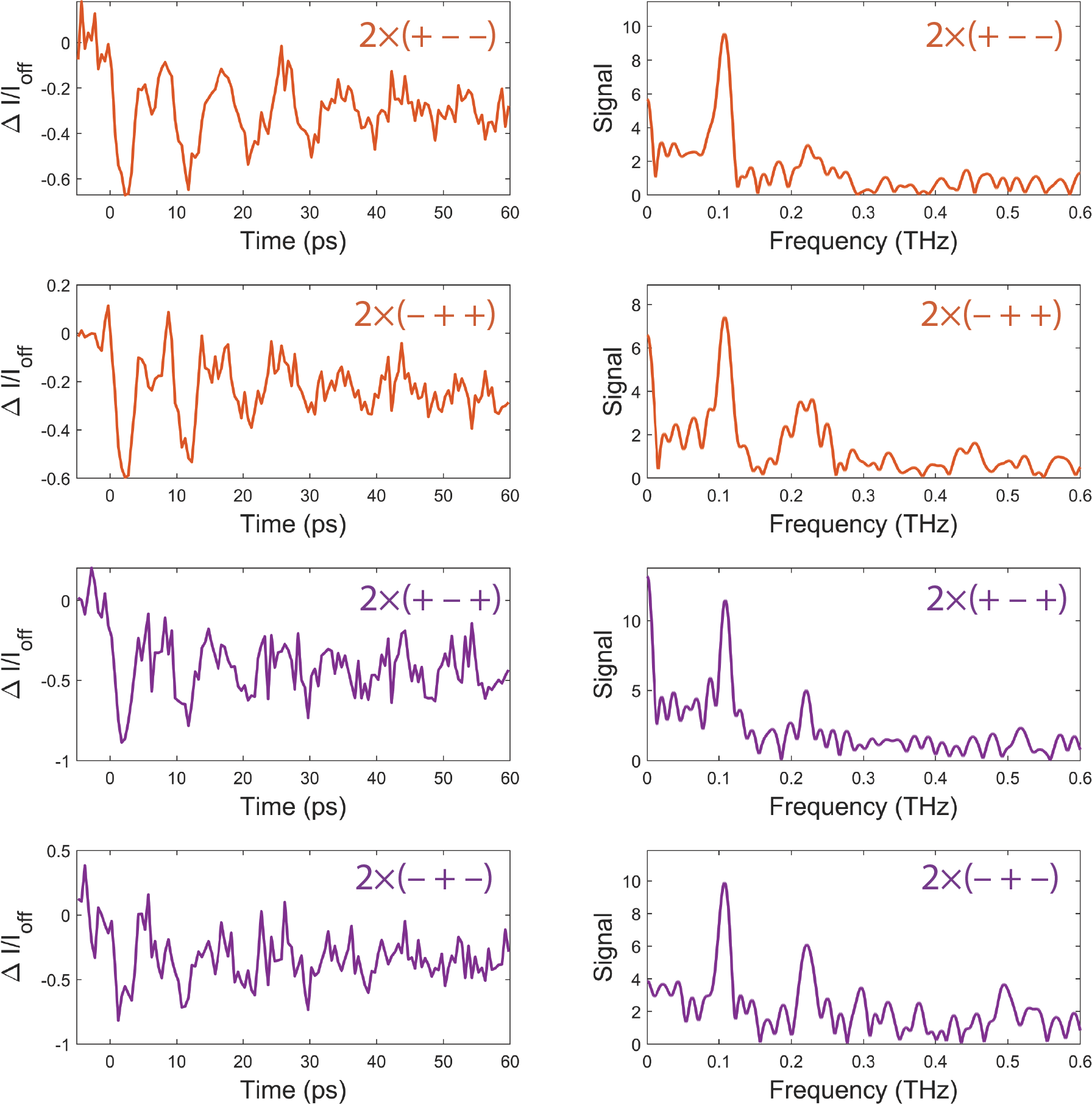}
    \caption{Data from the four observed second-order sidebands around the (2 2 4) crystal Bragg peak of TSI.}
    \label{fig:secondorder_all}
\end{figure}

The time domain signals ($\Delta I/I_{\textrm{off}}$, where $\Delta I(t) = I(t) - I_{\textrm{off}}$) and the Fourier spectra for all twelve observed CDW sidebands around the (2 2 4) crystal Bragg peak are shown in Figs. \ref{fig:firstorder_all} and \ref{fig:secondorder_all}. These measurements were obtained during the SACLA experiment described in the main text. The curves in Figs. \ref{fig:firstorder_all} and \ref{fig:secondorder_all} are color coded such that sidebands with the same color belong to the same orientational domain according to the structure determination of the CDW phase \cite{vanSmaalen2001}.

\providecommand{\noopsort}[1]{}\providecommand{\singleletter}[1]{#1}%